\documentclass[aps,prl,twocolumn,nobalancelastpage,showpacs,superscriptaddress]{revtex4}
\usepackage{bm}
\usepackage{graphicx}
\usepackage{amssymb}
\usepackage{color}

\begin{document}

\preprint{}

\title{Electron Correlation Driven Heavy-Fermion Formation in LiV$_2$O$_4$}

\author{P. E. J\"{o}nsson}
\altaffiliation[Current address: ]{Department of Physics, Uppsala University,  Box 530, SE-751 21 Uppsala, Sweden}
\affiliation{RIKEN, 2-1 Hirosawa, Wako,
Saitama 351-0198, Japan}
\affiliation{CREST, Japan Science and Technology Agency (JST), Kawaguchi, Saitama 332-0012, Japan}
\author{K. Takenaka}
\affiliation{RIKEN, 2-1 Hirosawa, Wako,
Saitama 351-0198, Japan}
\affiliation{CREST, Japan Science and Technology Agency (JST), Kawaguchi, Saitama 332-0012, Japan}
\author{S. Niitaka}
\affiliation{RIKEN, 2-1 Hirosawa, Wako,
Saitama 351-0198, Japan}
\affiliation{CREST, Japan Science and Technology Agency (JST), Kawaguchi, Saitama 332-0012, Japan}
\author{T. Sasagawa}
\affiliation{CREST, Japan Science and Technology Agency (JST), Kawaguchi, Saitama 332-0012, Japan}
\affiliation{Department of Advanced Materials Science, University of Tokyo, 5-1-5 Kashiwanoha, Kashiwa, Chiba 277-8581, Japan}
\author{S. Sugai}
\affiliation{Department of Physics, Nagoya University, Furo-cho, Chikusa-ku,
Nagoya 464-8602, Japan}
\author{H. Takagi}
\affiliation{RIKEN, 2-1 Hirosawa, Wako,
Saitama 351-0198, Japan}
\affiliation{CREST, Japan Science and Technology Agency (JST), Kawaguchi, Saitama 332-0012, Japan}
\affiliation{Department of Advanced Materials Science, University of Tokyo, 5-1-5 Kashiwanoha, Kashiwa, Chiba 277-8581, Japan}
\date{\today}

\begin{abstract}

Optical reflectivity measurements were performed on a single crystal
of the $d$-electron heavy-fermion (HF) metal LiV$_2$O$_4$. 
The results evidence the highly incoherent character of the charge dynamics
for all temperatures above $T^* \approx$ 20~K.
The spectral weight of the optical conductivity is redistributed over extremely broad energy scales ($\sim 5$~ eV) as the quantum coherence of the charge carriers is recovered. This wide redistribution is, in sharp contrast to $f$-electron Kondo lattice HF systems, characteristic of a metallic system close to a correlation driven insulating state.
Our results thus reveal that strong electronic correlation effects
dominate the low-energy charge dynamics and heavy quasiparticle formation in LiV$_2$O$_4$.
We propose the geometrical frustration, which limits the extension of charge and spin ordering,  as an additional key ingredient of  the low-temperature heavy-fermion formation in this system. 

\end{abstract}


\maketitle

Electrons in solids, by coupling with spins and lattices, form dressed particles called {\em quasiparticles} (QP). The mass of such QPs can in some cases be extremely heavy, 100-1000 times the bare electron mass.
The systems with extremely heavy QPs are called {\em heavy fermions} (HF) and has been attracting considerable interest since they display a variety of novel phenomena including exotic superconductivity \cite{degiorgi99}. 
Recently, a few
$d$-electron systems have been found to exhibit physical properties
characteristic of HF systems, e.g Y(Sc)Mn$_2$\cite{shietal93}, 
Na$_{0.75}$CoO$_2$ \cite{miyoshi2004} and LiV$_2$O$_4$
\cite{kondo97,krimmel99,uranoetal2000,shimoyamada2006}.
Conventional HF metals are $f$-electron systems containing rare-earth or actinide
ions, in which the low temperature heavy QP formation can be understood based on the Kondo-coupling between localized $f$-electron moments and itinerant electrons. 
In the case of  $d$-electron metals, it is not obvious to identify the same Kondo coupled itinerant and localized electrons.
A common point for the $d$-electron HF metals is instead that the magnetic ions occupy sites on geometrically frustrated lattices.
It has been discussed that the HF behaviour of $d$-electron systems might imply a new route to the formation of heavy quasiparticles. 

Among the $d$-electron HF metals  the spinel LiV$_2$O$_4$, has the largest quasiparticle specific heat coefficient $\gamma = 420$ mJ/mol.K$^2$ \cite{kondo97}.
It exhibits a short-range antiferromagnetic
order below 80~K \cite{leeetal2001}, and no other kind of long-range magnetic
order at any measured temperature. At a characteristic temperature, 
$T^* \approx 20$~K, the heat capacity over temperature $C/T$ shows a steep
increase and the electrical resistivity $\rho$ a sharp drop, suggesting that
coherence is formed \cite{uranoetal2000}. This, in addition to a
Fermi liquid ground state \cite{uranoetal2000} and a peak in the
density-of-states (DOS) located $\sim 4$~meV above the Fermi level
($E_{\rm F}$) \cite{shimoyamada2006}, confirm the low-$T$ HF properties.
Band structure calculations on LiV$_2$O$_4$  
indicate that mainly vanadium 3$d$ $t_{2g}$
bands crosses the Fermi level \cite{matsuno99,anietal99}. 
These V($3d$) $t_{2g}$ bands are well
separated from the filled O($2p$) bands and the empty V($3d$) $e_g$
bands. 
The triply degenerate $t_{2g}$ orbitals are splitted into doubly
degenerate $E_g$  and nondegenerate
$A_{1g}$  orbitals due to the trigonal crystal field. 
As a result of this trigonal splitting and $d$-$d$ Coulomb interaction,
it has been proposed
that the  $A_{1g}$ orbitals can be considered as localized and the
$E_g$ orbitals as itinerant permitting to map the electronic structure
of LiV$_2$O$_4$ into a Kondo-lattice picture \cite{anietal99}. 
It requires, however, a subtle cancellation of ferromagnetic coupling through double exchange with  antiferromagnetic coupling through Kondo-like exchange.
Other scenarios
have been proposed in which the geometrical frustration is an important
ingredient for the HF formation \cite{fulde2001,hopkinson2002}. Still, the
origin of the HF formation in LiV$_2$O$_4$ is a matter of controversy, which
can only be resolved by new detailed experiments. In particular, a method
probing the electronic structure over wide energy scales, such as optical
spectroscopy, may reveal valuable information about the mechanisms responsible for the low energy HF formation.

In this Letter, we report the first detailed investigation of the optical
properties of single crystals of LiV$_2$O$_4$.
The incoherent charge dynamics at $T>T^*$ and the transfer of spectral weight over broad energy scales ($\sim 5$~eV) reveal that LiV$_2$O$_4$, in contrast to conventional $f$-electron HF metals, is a correlated metal in proximity to a correlation driven insulating state. 
We outline a scenario in which strong correlation effects are controlling the formation of heavy QP states, while the key role of the geometrical frustration is to limit the extension of charge and spin orderings. 

Single crystals of LiV$_2$O$_4$ were grown by the flux method described
in Ref.~\cite{matuedued2005}. Near-normal incident reflectivity spectra
were measured on the as-grown shiny surface of crystals with octahedral
shape and \{111\} faces with edges of at most 1~mm. We confirmed that the
reflectivity at room temperature was the same as that of a small cleaved
surface. The reflectivity was measured using a Fourier-type interferometer
for the photon energy range of $\hbar \omega=$0.01--1.0~eV and grating
spectrometers for the energy range 0.5--6~eV. The crystal size was
sufficient for optical measurements using microscopes designed for the
infrared (IR) and visible-ultraviolet spectrometers. As a reference mirror
we used an evaporated Au film on a glass plate with the same shape and
size as the sample for the far-to-near IR regions, in order to cancel out
diffraction effects.  
The experimental error of the reflectivity, determined by the reproducibility, was less than 1\%.
The dc resistivity $\rho$ was measured using a four-probe technique.

\begin{figure}[tbh]
\includegraphics[width=\columnwidth]{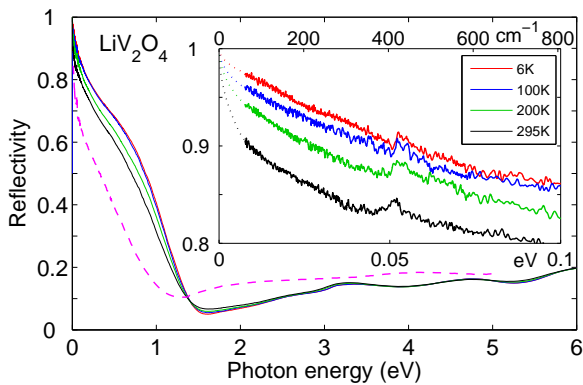}
\caption{\label{refl}
(color online). Optical reflectivity spectra measured on the as-grown surface of
LiV$_2$O$_4$ crystals at temperatures of 6, 100, 200 and 295~K (solid lines, from top to bottom) and on the polished surface (using lapping films with diamond powder of diameter 1$\mu$m)
at 295~K (dashed line).
The inset shows the low-energy part of the as-grown surface 
spectra with an optical phonon peak at
$\sim$420~cm$^{-1}$ (0.05~eV). The dotted lines are the extrapolations using the
Hagen-Rubens formula.}
\end{figure}

The reflectivity spectra measured on the 
as-grown surface at different temperatures are shown in
Fig.~\ref{refl}. A reflectivity edge, observed around 1.5~eV, reflects the
metallic character of this system, which produces the far-infrared (FIR)
spectral weight in the optical conductivity due to charge carriers. 
In ordinary metals, the edge is steep and the reflectivity below
the edge is already close to 1. However, in LiV$_2$O$_4$ the reflectivity
shows a gradual increase below the edge and the reflectivity in the
mid-infrared (MIR) and near-infrared (NIR) regions are rather suppressed, 
suggesting the incoherent nature of the charge carriers. 
With decreasing temperature the reflectivity edge
becomes sharper, and the reflectivity in the FIR to MIR region higher, 
indicating that coherence is gained at lower temperatures. No significant
change was observed in the reflectivity spectra from 50~K down to the
lowest measured temperature of 6~K in the considered frequency range.

The room
temperature reflectivity for a polished surface is also shown
in Fig.~\ref{refl}. 
The reflectivity of the polished surface is significantly reduced in the
mid-to-near IR region and rather enhanced in the visible-to-ultraviolet 
region. 
If the scattering of the incident light is increased simply 
due to residual surface roughness, 
the reflectivity measured on the polished surface would be suppressed over
{\it all} energy regions. 
We therefore consider that the observed variation of the 
reflectivity reflects changes in the electronic states. 
The charge excitations are sensitive to
static imperfections and/or structural strain. 
The suppressed IR
reflectivity indicates the tendency of the charge carriers to localize while the enhanced
visible-to-ultraviolet reflectivity suggests an intimate relation between
the charge carrier dynamics  and the higher-lying interband
transitions. 
This sensitivity of the reflectivity to surface treatment,
similar to that of some manganites in a metallic phase \cite{takenaka99}, 
 suggests the importance of electronic correlations in LiV$_2$O$_4$ \cite{millis03}.

Group theory predicts that four $F_{1u}$ phonons are IR active in a cubic
spinel. For insulating cubic spinels, the two lower-energy modes
have much smaller spectral weight compared with the two
higher-energy modes \cite{lutz91,sushkov05}. 
In the high electrical resistivity metal or semiconducting MgTi$_2$O$_4$, only the two higher-energy phonons  are observed  at 420~cm$^{-1}$ and 600~cm$^{-1}$ \cite{popovic2003}. 
The 420~cm$^{-1}$ mode having the largest spectral weight in the metallic (semiconducting) phase.
The absence of the two lower-energy modes is ascribed to the screening by free carriers.
Only one phonon is observed at
420 cm$^{-1}$ in metallic LiV$_2$O$_4$ (inset of Fig.~\ref{refl}). 
Taking into account
the screening effect, the other phonon modes are likely buried in the noise level of the reflectivity spectrum.

\begin{figure}[tbh]
\includegraphics[width=\columnwidth]{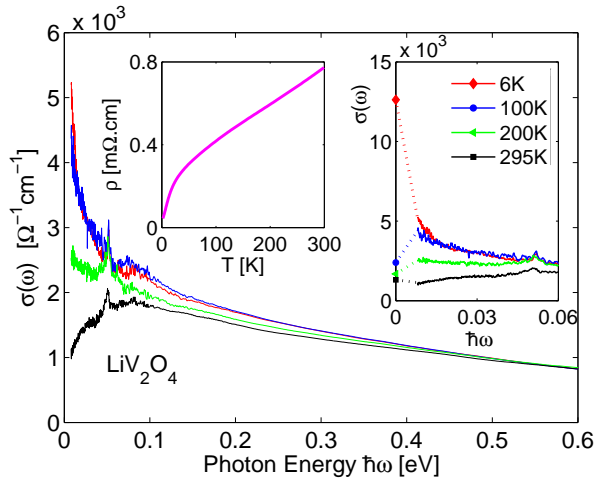}
\caption{\label{sigma_lowE}
(color online). Real part of $\sigma(\omega)$ of LiV$_2$O$_4$ at $T$=6, 100, 200 and 
295~K (from top to bottom). 
The peaks near 0.05~eV are due to optical phonons. 
The left inset shows the  temperature dependence of the resistivity ($\rho = \sigma_{\rm dc}^{-1}$).
The right inset the connection between $\sigma(\omega)$ (solid lines) and $\sigma_{\rm dc}$ (markers), the dotted lines are only guides for the eye.
}
\end{figure}

The optical conductivity $\sigma(\omega)$ was determined from the reflectivity data 
by Kramers-Kronig
transformation. To do this, we assumed the Hagen-Rubens formula in the
low-energy region and used a constant reflectivity above 6~eV 
followed by a well-known function of $\omega^{-4}$ in the vacuum-ultraviolet region. 
The optical conductivity in the low-energy region is shown 
in Fig.~\ref{sigma_lowE}, and the connection with the 
dc conductivity $\sigma_{\rm dc}(=\rho^{-1})$ in the right inset.

The evolution of $\sigma(\omega)$ as a function of temperature clearly demonstrates a crossover from coherent to incoherent charge dynamics. At $T=6$~K (below $T^* \approx 20$~K), by extrapolating the measured $\sigma(\omega)$ to $\sigma_{\rm dc}$, a clear Drude contribution is identified at low energies, consistent with the formation of coherent quasiparticles.
The Drude contribution shows a very slow decay with $\omega$, which is typical for strongly correlated transition-metal oxides.
A sharp coherence peak in the DOS, located $\sim 4$~meV above $E_{\rm F}$, was observed recently by high-resolution photoemission spectroscopy at temperatures below $T^*$
\cite{shimoyamada2006}. 
The DOS of $f$-electron HF systems exhibits a similar low-$T$ DOS peak, known as the Kondo resonance,  yielding an 
narrow Drude peak in $\sigma(\omega)$ \cite{degiorgi99}.
In addition, a pseudogap  and a broad MIR peak, originating from optical inter-band transitions between renormalized hybridization bands,  appears for many $f$-electron HF systems \cite{dordevic2001,degiorgi2001}.
While the low temperature narrow Drude peak is observed in $\sigma(\omega)$ of   LiV$_2$O$_4$, the broad MIR peak and pseudogap cannot clearly be identified.

Above $T^*\approx 20$~K, the Drude contribution is not well defined anymore, instead, a finite energy peak shows up in $\sigma(\omega)$. This is most clearly seen in the spectrum at 295~K, where a broad peak centered around 0.1~eV is observed.
The presence of such finite energy peak is less clear at lower temperatures. However, the fact that $\sigma_{\rm dc}$ above $T^*\approx 20$~K is much lower than naively anticipated from low energy $\sigma(\omega)$ clearly indicates the presence of a low energy peak.
The resistivity, shown in the left inset of Fig.~\ref{sigma_lowE},  exhibits almost $T$-linear dependence above $T^*$, followed by a rapid decrease below $T^*$ with decreasing temperature. Although the temperature dependence of $\rho$ is seemingly metallic above $T^*$, the large magnitude of the resistivity in this temperature range yields $k_F l\lesssim 1$ \cite{uranoetal2000}.
Combining this violation of the Ioffe-Regel limit and the absence of a Drude peak, it is clear that the system is incoherent above $T^*$.

\begin{figure}[tbh]
\includegraphics[width=\columnwidth]{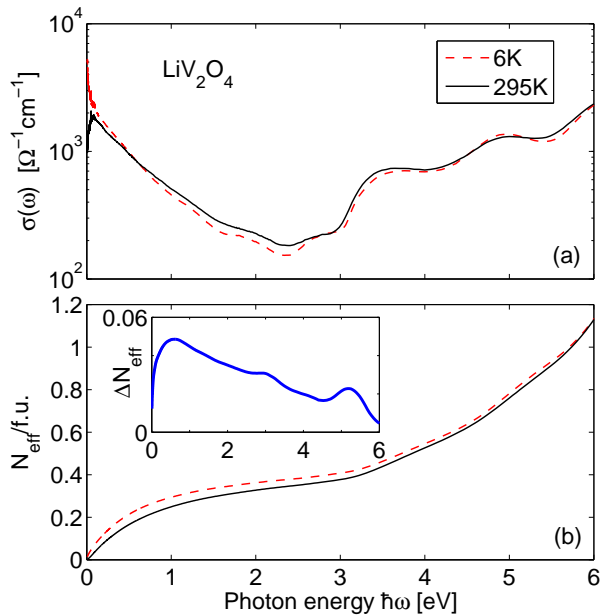}
\caption{\label{fig-neff}
(color online). (a) Real part of $\sigma(\omega)$ on a semilogarithmic scale.
(b)
Integrated spectral weight $N_{\rm eff}$ per formula unit. 
Inset:
$\Delta N_{\rm eff}(\omega) =\frac{2m_0V}{\pi e^2}\int_{0}^{\omega}[\sigma_{6K}(\omega')-\sigma_{295K}(\omega')]d\omega' $.
}
\end{figure}

With increasing
temperature to above $T^*$, the FIR conductivity does not only become incoherent but also
looses its spectral weight. The missing spectral weight is transferred over
an extremely broad energy region as shown in Fig.~\ref{fig-neff}. 
The effective carrier density defined as 
\begin{equation}
N_{\rm eff} =\frac{2m_0V}{\pi e^2}\int_{0}^{\omega}\sigma(\omega')d\omega' 
\end{equation}
(with $m_0$ being the bare electron mass and $V$ the unit-cell volume)
indicates that the $f$-sum rule is  finally fulfilled above $\sim 5$~eV. 
Hence, the recovery of quantum coherence is accompanied by a spectral
weight transfer over energies larger than the bandwidth $W$ of the conduction band. 
This is a feature characteristic of correlated metallic
systems close to a Mott insulating or charge ordered state \cite{imada98}, 
but in sharp contrast to $f$-electron  HF systems in the Kondo-lattice regime\cite{degiorgi99,logan2005}.
In HF metals, spectral weight redistribution is confined to energies $\hbar \omega < W$ \cite{hussey2004}.
Typically this energy scale is very low of the order of 10-100~meV \cite{degiorgi99,logan2005}.
The possibility of Kondo lattice formation  resulting from the band structure and $d-d$ Coulomb correlations has been proposed for LiV$_2$O$_4$ \cite{anietal99}.
In that model, the mechanism for the heavy QP formation is similar as for $f$-electron systems, i.e. hybridization between the ``localized'' $A_{1g}$ band and the ``itinerant'' $E_g$ band.
While such a model may explain the resonance peak in the DOS,  it is not obvious that  it can account for the redistribution of spectral weight over the energy range of 5~eV in Fig.~\ref{fig-neff}.

The incoherent transport in LiV$_2$O$_4$ is similar to that of a 
wide range of correlated bad
metals, including vanadates \cite{rozenberg95}, manganites
\cite{takenaka2002}, cobaltates \cite{wang04}, cuprates \cite{takenaka2003}, 
ruthenates \cite{kostic98}, and organic conductors \cite{takenaka2005}.
In correlated metals, the
coherence temperature is suppressed down to room temperature or lower and
the charge dynamics is incoherent at higher temperatures. In the incoherent regime,  
$\sigma(\omega)$ is characterized by a finite-energy peak, instead of a
conventional zero-energy (Drude) peak. The  metallic $T$-dependence of
the dc resistivity does not reflect the broadening of a Drude peak as in ordinary metals, but the collapse of the Drude peak into a finite energy peak and the reduction of the FIR spectral weight with increasing temperature. 
These features typical of correlated bad metals, as well as
the transfer of spectral weight over a large energy scale (shown in Fig.~\ref{fig-neff}) demonstrate that strong electronic
correlations dominate the charge dynamics and the recovery of the quantum coherence
at low temperatures in LiV$_2$O$_4$.

LiV$_2$O$_4$ is a mixed valent spinel with equal ratio of V$^{3+}$ and V$^{4+}$, and it is very likely located in the close vicinity of a charge ordered state.
In general, the QP mass is enhanced when approaching a correlation driven metal-insulator transition from the metallic side \cite{tokura93miyasaka2000}. 
In most cases, however, spin and/or charge ordering occurs before any gigantic
enhancement of the mass is realized, giving rise to only a moderately enhanced mass. 
In LiV$_2$O$_4$ however, 
the long range spin-orbital-charge order is prevented by the geometrical frustration and hence the mass
enhancement may proceed further. 
Accordingly, theoretical studies of simple models for strongly correlated electron systems suggest a considerable
suppression of  the temperature scale of electronic coherence by magnetic frustration
\cite{parcollet99} and 
a Kondo-like resonance in the metallic phase close to a Mott or charge-ordered transition \cite{ohashi}.

In summary, we have investigated the temperature dependent optical
properties of LiV$_2$O$_4$ using well-characterized single crystals. A
strongly incoherent charge dynamics is found for all temperatures above
$T^* \approx 20$~K, in agreement with the ``bad-metal'' behavior observed
in resistivity measurements. The transfer of spectral weight occurs over
an extremely wide energy range  when the quantum coherence of the electrons
is recovered, as observed in correlated metals 
close to a correlation driven insulating state.  
This clearly indicates that strong correlation effects are controlling the formation of quasiparticle states at low energies in LiV$_2$O$_4$. 
We propose the geometrical frustration, which limits the extension of charge and spin ordering,  as an additional  key ingredient of  the low-temperature HF formation.
This general scenario might be extended to other geometrical frustrated $d$-electron metals with heavy quasiparticles.

We are grateful to K. Okazaki for his help in the IR measurements and
to J. Matsuno for valuable discussions. This work was supported by 
Grant-in-Aids for Creative Scientific Research from the Ministry of
Education, Culture, Sports, Science and Technology of Japan.


\begin{thebibliography}{10}

\bibitem{degiorgi99}
L. Degiorgi, Rev. Mod. Phys. {\bf 71},  687  (1999).

\bibitem{shietal93}
M. Shiga, K. Fujisawa, and H. Wada, J. Phys. Soc. Jpn. {\bf 62},  1329  (1993).

\bibitem{miyoshi2004}
K. Miyoshi, E. Morikuni, K. Fujiwara, J. Takeuchi, and T. Hamasaki, Phys. Rev.
  B {\bf 69},  132412  (2004).

\bibitem{kondo97}
S. Kondo {\it et al.}, Phys. Rev. Lett. {\bf 78},  3729  (1997).

\bibitem{krimmel99} A. Krimmel, A. Loidl, M. Klemm, S. Horn, and H. Schober, 
Phys. Rev. Lett. {\bf 82}, 2919 (1999).

\bibitem{uranoetal2000}
C. Urano, M. Nohara, S. Kondo, F. Sakai, H. Takagi, T. Shiraki, and T. Okubo,
  Phys. Rev. Lett. {\bf 85},  1052  (2000).

\bibitem{shimoyamada2006}
A. Shimoyamada {\it et al.},
Phys. Rev. Lett. {\bf 96},  026403  (2006).


\bibitem{leeetal2001}
S.-H. Lee, Y. Qiu, C. Broholm, Y. Ueda, and J.~J. Rush, Phys. Rev. Lett {\bf
  86},  5554  (2001).


\bibitem{matsuno99}
J. Matsuno, A. Fujimori, and L.~F. Mattheiss, Phys. Rev. B {\bf 60},  1607
  (1999).

\bibitem{anietal99}
V.~I. Anisimov, M.~A. Korotin, M. Z{\"o}lfl, T. Pruschke, K.~LeHur, and T.~M.
  Rice, Phys. Rev. Lett {\bf 83},  364  (1999).

\bibitem{fulde2001}
P. Fulde, A.~N. Yaresko, A.~A. Zvyagin, and Y. Grin, Europhys. Lett. {\bf 54},
  779  (2001).

\bibitem{hopkinson2002}
J. Hopkinson and P. Coleman, Phys. Rev. Lett. {\bf 89},  267201  (2002).


\bibitem{matuedued2005}
Y. Matsushita, H. Ueda, and Y. Ueda, Nature Materials {\bf 4},  845  (2005).


\bibitem{takenaka99}
K. Takenaka, K. Iida, Y. Sawaki, S. Sugai, Y. Moritomo, and A. Nakamura, 
J. Phys. Soc. Jpn. {\bf 68}, 1828 (1999).

\bibitem{millis03} A. J. Millis, Solid State Commun. {\bf 126}, 3 (2003).

\bibitem{lutz91}
H.~D. Lutz, B. M{\"u}ller, and H.~J. Steiner, J. Solid State Chem. {\bf 90},
  54  (1991).


\bibitem{sushkov05}
A. B. Sushkov, O. Tchernyshyov, W. Ratcliff , S. W. Cheong, and H. D. Drew, 
Phys. Rev. Lett. {\bf 94}, 137202 (2005).

\bibitem{popovic2003}
Z. V. Popovic, G.  De Marzi, M. J. Konstantinovi\'{c}, A. Cantarero, Z. Dohcevic-Mitrovic, M.  Isobe,  and Y. Ueda,
Phys. Rev. B {\bf 68}, 224302 (2003).


\bibitem{dordevic2001}
S. V. Dordevic, D. N Basov, N. R.  Dilley, E. D. Bauer, and M. B. Maple, 
Phys. Rev. Lett., {\bf  86}, 684 (2001).

\bibitem{degiorgi2001}
L. Degiorgi, F. B. B. Anders, and G. Gr{\"u}ner, Eur. Phys. J. B {\bf 19}, 167 (2001).

\bibitem{imada98}
O. Gunnarsson, M. Calandra, and J. E. Han, Rev. Mod. Phys. {\bf 75}, 1085 (2003);
M. Imada, A. Fujimori, and Y. Tokura, Rev. Mod. Phys. {\bf 70},  1039  (1998).

\bibitem{logan2005}
D. E. Logan and N. S. Vidhyadhiraja, J. Phys.:Condens. Matter {\bf 17}, 2935 (2005);
N. S. Vidhyadhiraja and D. E. Logan, J. Phys.:Condens. Matter {\bf 17}, 2959 (2005). 


\bibitem{hussey2004}
N. E. Hussey, K. Takenaka, and H. Takagi, Phil. Mag. {\bf 84}, 2847 (2004).


\bibitem{rozenberg95}
M.~J. Rozenberg, G. Kotliar, H. Kajueter, G.~A. Thomas, D.~H. Rapkine, J.~M.
  Honig, and P. Metcalf, Phys. Rev. Lett. {\bf 75},  105  (1995).

\bibitem{takenaka2002}
K. Takenaka, Y. Sawaki, and S. Sugai, Phys. Rev. B {\bf 60}, 13011 (1999); 
K. Takenaka, R. Shiozaki, and S. Sugai, Phys. Rev. B
{\bf 65}, 184436 (2002).

\bibitem{wang04}
N. L. Wang, P. Zheng, D. Wu, Y. C. Ma, T. Xiang, R. Y. Jin, and D. Mandrus, 
Phys. Rev. Lett. {\bf 93}, 237007 (2004).

\bibitem{takenaka2003}
K. Takenaka, J. Nohara, R. Shiozaki, and S. Sugai, Phys. Rev. B {\bf 68}, 
134501 (2003); D. N. Basov and T. Timusk, Rev. Mod. Phys. {\bf 77}, 721
(2005).

\bibitem{kostic98}
Y. S. Lee, Jaejun Yu, J. S. Lee, T. W. Noh, T.-H. Gimm, 
Han-Yong Choi, and C. B. Eom, Phys. Rev. B {\bf 66}, 041104(R) (2002).

\bibitem{takenaka2005}
K. Takenaka, M. Tamura, N. Tajima, H. Takagi, J. Nohara, and S. Sugai, 
Phys. Rev. Lett. {\bf 95}, 227801 (2005).

\bibitem{tokura93miyasaka2000}
Y. Tokura, Y. Taguchi, Y. Okada, Y. Fujishima, T. Arima, K. Kumagai, and Y. Iye, Phys. Rev. Lett. {\bf 70}, 2126 (1993).

\bibitem{parcollet99} O. Parcollet and A. Georges, Phys. Rev. B {\bf 59}, 
5341 (1999); J. Merino and R. H. McKenzie, Phys. Rev. B {\bf 61}, 7996 (2000).

\bibitem{ohashi}
T. Ohashi, N. Kawakami, and H. Tsunetsugu, Phys. Rev. Lett. {\bf 97}, 066401 (2006);
H. Kusunose, S.  Yotsuhashi, and K. Miyake,
Phys. Rev. B, {\bf 62}, 4403 (2000).

\end{thebibliography}
\end{document}